\documentstyle[preprint,aps,floats]{revtex}
\tightenlines
\input{epsf.tex}
\begin{document}
%\rightline{DAMTP-1999-50}
\def\ba{\begin{eqnarray}}
\def\ea{\end{eqnarray}}
\def\be{\begin{equation}}
\def\ee{\end{equation}}
%\def\tr{{\rm tr}}
%\def\sech{{\rm sech }}
%\def\gtorder{\mathrel{\raise.3ex\hbox{$>$}\mkern-14mu
%             \lower0.6ex\hbox{$\sim$}}}
%\def\ltorder{\mathrel{\raise.3ex\hbox{$<$}\mkern-14mu
%             \lower0.6ex\hbox{$\sim$}}}

%\title{Gravitational effects in the Colliding Bubble Braneworld Scenario}

\title{When do colliding bubbles produce an expanding universe?}

\author{Jose J. Blanco-Pillado$^{1,}$\thanks{
E-mail: J.J.Blanco-Pillado@damtp.cam.ac.uk}, 
Martin Bucher$^{1,}$\thanks{E-mail: M.A.Bucher@damtp.cam.ac.uk},
Sima Ghassemi$^{1,2,}$\thanks{E-mail: sima\_gh@mehr.sharif.ac.ir}
and Frederic Glanois$^{1}$.\vspace{.3cm}\\}
\address{$^{1}$
DAMTP, Centre for Mathematical Sciences,\\
University of Cambridge, Wilberforce Road,\\
Cambridge, CB3 0WA\\
United Kingdom\vspace{.3cm}\\}
\address{$^{2}$
Department of Physics, Sharif University of Technology,\\
Tehran, P.O.Box: 11365-9161, Iran\vspace{.2cm}\\}

\date{June 2003}
 
%\date{January 2003} 
%\date{February 2003} 
%\date{April 2003} 

\maketitle

\begin{abstract}
{
It is intriguing to consider the possibility that the Big Bang of
the standard (3+1) dimensional cosmology originated from the collision
of two branes within a higher dimensional spacetime, leading to the 
production of a large amount of entropy. In this paper we study,
subject to certain well-defined assumptions, under what conditions such a 
collision leads to an expanding universe. We assume the absence of novel
physics, so that ordinary (4+1)-dimensional Einstein gravity remains a 
valid approximation. It is necessary that the fifth dimension not become degenerate
at the moment of collision. First the case of a symmetric collision of
infinitely thin branes having a hyperbolic or flat spatial geometry is
considered. We find that a symmetric collision results in a collapsing
universe on the final brane unless the pre-existing 
expansion rate in the bulk just prior to the collision is sufficiently
large in comparison to the momentum transfer in the fifth dimension.
 Such prior expansion may either result from negative spatial
curvature or from a positive five-dimensional
cosmological constant. The relevance of these findings to the Colliding
Bubble Braneworld Universe scenario is discussed. Finally, results
from a numerical study of colliding thick-wall branes is presented,
which confirm the results of the thin-wall approximation. 
}
\end{abstract}
\newpage
\section{Introduction.}

\bigskip
Recently much work motivated by ideas from string/M-theory has
been devoted to the study of large extra dimensions\cite{ADD,rs}. Much 
of this work has been done within the so-called braneworld scenarios, where all
the low-energy excitations except for gravity are confined to a brane, embedded 
in the simplest case in five-dimensional anti de Sitter space ($AdS^{5}$)\cite{rs}.
It has been shown that in this type of model, gravity behaves in the large distance
limit as ordinary $(3+1)$ dimensional gravity\cite{rs,gt1}.
It has also been shown that in the limit where the energy density on the brane is small,
the standard four-dimensional cosmological evolution on the brane 
is recovered with calculable corrections that become relevant at high densities\cite{langloisa}. 

Given this starting point, there exist a number of ways to proceed toward constructing
a more complete cosmological model. Perhaps the most conservative
route is to postulate an epoch of inflationary expansion on the brane. However,
when the extra dimensions are large, it is not at all clear that such inflation
can account for the high degree of smoothness of our present universe. A
(4+1)-dimensional (``bulk") smoothness problem arises\cite{bucher,trodden}.
Unless some additional mechanism exists for imposing smoothness on the five-dimensional 
bulk spacetime,             
after inflation on the brane ends, its subsequent evolution into an irregular
bulk is likely to produce unacceptably large irregularities on the brane.
Consequently, considerable motivation exists for exploring possibilities
for a more complete story for a braneworld cosmology,
possibly not involving inflation at all.

A number of models, in particular brane inflation\cite{dvali}, the 
ekpyrotic universe\cite{ekpyrotic}, and the colliding
bubble braneworld scenario\cite{bucher}, postulate that the large entropy of the present
universe has its origin in the collision of branes. In the latter scenario
\cite{bucher,jjmb1,gt2,jjmb2} two expanding bubbles collide to form a final 
brane on which our (3+1)-dimensional FRW universe is situated. 
In this model, the universe starts in a metastable state with a 
(4+1)-dimensional Minkowski or de Sitter geometry. In the former case
a natural preferred ground state exists. In the latter case, an epoch
of old inflation \`a la Guth rids the spacetime of any pre-existing
irregularities, effectively establishing an $SO(5,1)$ symmetric initial
state with calculable fluctuations departing from this symmetry. 
This initial state decays through quantum tunneling 
by forming bubbles filled with anti 
de Sitter space\cite{fvd,deluccia}.
If the true vacuum is discretely degenerate, the collision of two such 
bubbles results in the formation of a domain wall 
interpolating between two different such vacua. This is our braneworld. 
If the tension of this local brane is tuned to balance the negative
cosmological constant in the bulk, the cosmology for an observer
situated on this brane is virtually identical to that for a 
Randall-Sundrum model. The decay of one bubble gives 
a spacetime with its symmetry reduced to $SO(4,1).$ 
For two colliding bubbles, the symmetry reduces further 
to $SO(3,1)$\cite{hms,wu}. On our braneworld, which has a 
three-dimensional hyperbolic geometry $H_{(3)},$ this is the symmetry group of 
spatial manifold at constant cosmic time. 
If the curvature radius today $R_0$ of this $H_{(3)}$ manifold is sufficiently large, the universe will appear very nearly flat. 

After the collision, the evolution of the FRW universe is governed by
the modified FRW equation given in ref.~\cite{langloisa}. Given the 
density on the final brane, which is fixed by stress-energy-momentum 
conservation, the future time evolution of this FRW universe is completely
determined. However, because the Hubble constant on the brane $H$
appears quadratically in this equation,
it is necessary to determine the sign of $H$ on the final brane. 

In this paper, we examine the outcome of symmetric brane collisions, first
in the infinitely thin brane limit and then numerically for the
collision of thick branes. Although this work 
was initially motivated by the Colliding Bubble Braneworld scenario, the 
calculations and techniques developed are more generally applicable. 

The organisation of the paper is as follows. In section II we calculate
the outcome of collisions of thin branes. We also indicate how these 
results would be modified when our assumptions are relaxed. We show
how the same conclusions may be obtained more generally by considering 
the focusing of geodesics as a result of the momentum transfer in the
 collision.
In sections III and IV we show how to simulate bubble collisions
numerically and present the results of our numerical simulations, which
agree with the analytic treatment in section II. Finally, in section V
we present some concluding remarks.

\section{Collisions of branes without thickness}

\begin{figure}
\begin{center}
\epsfxsize=3in
\epsfysize=3in
\begin{picture}(300,200)
\put(145,220){$B_F$}
\put(70,30){$B_L$}
\put(215,30){$B_R$}
\put(145,40){$M^5$}
\put(70,140){$AdS^5$}
\put(215,140){$AdS^5$}
\put(285,70){time}
\put(95,-12){fifth dimension}
\put(80,1){\leavevmode\epsfbox{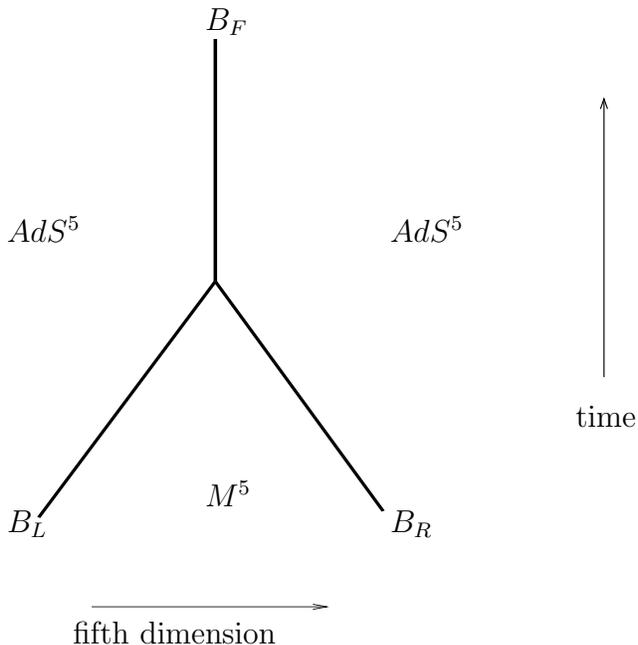}}
\end{picture}
\end{center}
\vskip 15pt
\caption{\small {{\bf Brane Collision Geometry.}
Two branes collide to form a single final brane 
surrounded by pristine
bulk spacetime. Here each point represents an
$H_{(3)}$ manifold of variable curvature radius $R$
comprising the three transverse dimensions.}}
\label{Fig1}
\end{figure}

In this section we calculate the outcome of the symmetric 
collision of two infinitely thin branes that join to form a single brane 
in the final state.  It is assumed that the resulting brane is
surrounded by $AdS^5$ and that no debris emanates from
the collision. Brane collisions in the thin-wall limit have been previously 
considered by a number of authors\cite{bucher,Neronov,B7,wands,Berezin}.
The brane collision geometry is sketched 
in Fig.~\ref{Fig1}, where only two of the (4+1) dimensions are shown.
Because of a generalization of Birkhoff's theorem, the outcome of the collision may
be analysed by pasting together across the branes the two exterior $AdS^5$ spacetimes 
and the $M^5$ spacetime initially between the two colliding branes
in such a way that $R$, the curvature radius of the transverse $H_{(3)}$
manifold, is continuous across the cuts and that the proper
time differences on the two sides of the brane agree. Equivalently, it suffices
to match $R$ at one point and to require that the ``Hubble
constant"
\be
H={~d\over d\tau }\ln [R]
\ee
agree on the two sides of the brane. 
Here $\tau $ denotes the proper time along the brane.
$H(\tau)$ is entirely intrinsic
to the brane and fixes the trajectories of the boundaries in the bulk
spacetimes that have been glued together across the branes. 
The matching at the collision vertex is fixed by the absence 
of a conical curvature singularity and by stress-energy conservation.
Note that the equation of state on the brane is necessary to determine
$\dot H$; however, for matching at a collision vertex as considered
here only knowledge of $H$ is required. 

\begin{figure}
\begin{center}
\epsfxsize=3in
\epsfysize=3in
\begin{picture}(300,200)
\put(140,-10){$R=0$}
\put(145,60){{\bf I}}
\put(145,155){{\bf I}}
\put(140,225){$R=0$}
\put(7,105){$R=\infty $}
\put(90,105){{\bf II}}
\put(190,105){{\bf II}}
\put(260,105){$R=\infty $}
\put(40,1){\leavevmode\epsfbox{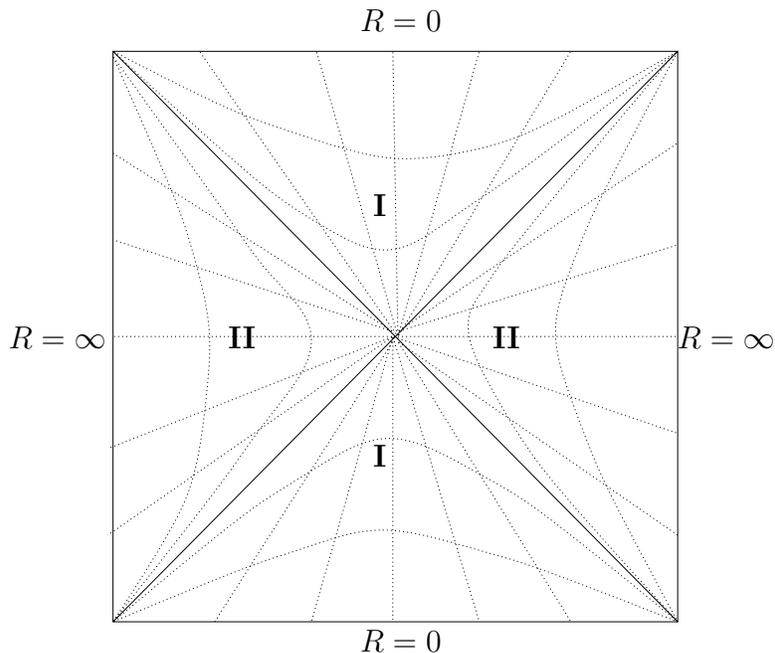
}}
\end{picture}
\end{center}
\caption{{\bf Hyperbolic coordinates for describing brane collisions in AdS.}
Squares such as the one illustrated above tile the infinite vertical strip comprising the
Penrose diagram for maximally extended $AdS$ space. Here four distinct
coordinate patches separated by diagonal coordinate singularities cover
the fundamental region.
}
\label{Fig2}
\end{figure}

In order to describe the surrounding $AdS^5$ bulk,
having a curvature radius $\ell ,$ it is necessary to employ
coordinates that make manifest the $H_{(3)}$ symmetry of the transverse dimensions.
There are two such coordinate systems, covering different 
patches of $AdS^5$ and separated by coordinate singularities. The type I
patches have the line element
\be 
ds^2=-dt^2+\cos ^2[t/\ell ]d\chi ^2+\ell ^2\sin ^2[t/\ell ]dH_{(3)}^2,
\ee
which in the $M^5$ $(\ell \to \infty )$ limit becomes
\be 
ds^2=-dT^2+dX^2+T^2dH_{(3)}^2.
\label{mink}
\ee
The type II patches have the line element
\be
ds^2=-\ell ^2\sinh ^2[\chi /\ell ]dt^2+d\chi ^2+\ell ^2\cosh ^2[\chi /\ell ]~dH_{(3)}^2,
\label{patchII}
\ee
which in the $M^5$ limit becomes the Rindler wedge
\be
ds^2=-\chi ^2dt^2+d\chi ^2+dE_{(3)}^2.
\ee
Here $E_{(3)}$ is simply flat three-dimensional Euclidean space.

\begin{figure}
\begin{center}
\epsfxsize=3in
\epsfysize=3in
\begin{picture}(300,200)
\put(145,-10){N}
%\put(40,1){\leavevmode\epsfbox{Fig3.eps
\put(40,1){\leavevmode\epsfbox{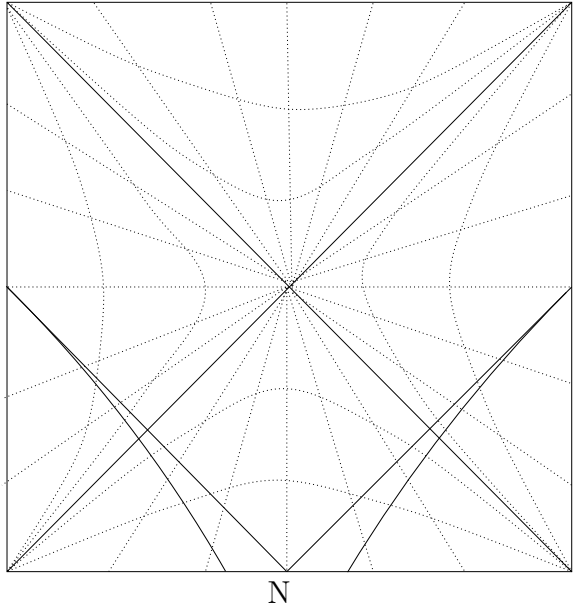
}}
\end{picture}
\end{center}
\caption{{\bf Expanding Bubble in Hyperbolic Coordinates.}
We show the placement of an expanding bubble with nucleation
center $N$ with respect to the above hyperbolic coordinates.
}
\label{Fig3}
\end{figure}

We now examine how the conformal diagram of maximally extended $AdS^5$ 
is covered by such coordinate patches. The conformal diagram consists
of an infinite vertical strip of finite width tiled by an infinite sequence
of square fundamental regions. Fig.~\ref{Fig2} shows how each of these fundamental
regions is covered by type I and type II coordinate patches. Note that
each point in this figure represents an $H_{(3)}$ manifold of variable curvature radius $R$
along the three transverse dimensions. The lower
horizontal boundary, with $R=0$, represents an infinite length spacelike 
geodesic $S.$ The region spanned by the coordinate patch of type I just above $S$
is generated by the family of all timelike geodesics emanating normally
from $S.$ These geodesics are focused in the horizontal direction to join at a point
situated in the middle
of the square, which represents an $H_{(3)}$ manifold with $R=\ell .$ The two
diagonal boundaries represent coordinate singularities along which
this patch joins two type II patches. Note that within the type I
patches, $0\le R<\ell .$ Inside the pair of type II patches, $\ell <R<+\infty .$
Fig.~\ref{Fig3} shows how an isolated expanding bubble is situated in relation to
these coordinates.
Since we are interested in bubbles that collide after much expansion
has taken place, we consider a collision in a type II patch, where $R>\ell .$

A collision may be characterized by two parameters: 
the curvature radius at collision $R_{coll},$ and the boost parameter
$\beta _{coll},$ defined such that the relative velocity of the colliding
branes is $v_{rel}=\tanh [2\beta _{coll}].$ Given $R_{coll}$ and $\beta _{coll},$
we may match, on the $M^5$ side with geometry (\ref{mink}), onto the trajectory
of an expanding bubble of the form,
\be
T=a~\sinh [\beta ],\quad \quad  X=a~\cosh [\beta ],
\ee
where
\be
a={R_{coll}\over \sinh [\beta _{coll}]}
\ee 
is the curvature radius of the $dS^4$ hyperboloid swept out. In terms of
the $H_{(3)}$ coordinates, the expansion rate of the three transverse dimensions
just before the collision is given by 
\be
H_{in}={~d\over d\tau }\ln [R]=
{1\over a}{~d\over d\beta }\ln \Bigl(a~\sinh [\beta ]\Bigr)
={1\over a}\coth [\beta ]={\cosh [\beta ]\over R}. 
\ee 
Therefore at the collision point we have
\be
H_{in}={\cosh [\beta_{coll}]\over R_{coll}}. 
\ee 
For the $AdS^5$ side of the bulk, we take the line element (\ref{patchII}) to obtain
the scale factor
\be
R=\ell \cosh [\chi /\ell ].
\ee 
It follows that $R_{coll}=\ell \cosh [\chi _{coll}/\ell ].$ The boost parameter with 
respect to the $AdS^5$ rest frame $\beta _{AdS}$ is related to $H$ by
\be
H=\sinh [\beta _{AdS}]{~d\over d\chi  }\ln \Bigl( \ell ~\cosh [\chi/\ell ]\Bigr)
={1\over \ell }\sinh [\beta _{AdS}]~\tanh [\chi /\ell ].
\ee 
Therefore, just before the collision
\be 
H_{in}={1\over \ell }\sinh [\beta _{AdS}^{in}]~\tanh [\chi _{coll}/\ell ].
\ee
It follows that
\be
\coth [\beta _{AdS}^{in}]=\sqrt{1+{R_{coll}^2-\ell ^2\over (\ell H_{in})^2~R_{coll}^2}}.
\ee

\begin{figure}
\begin{center}
\epsfxsize=3in
\epsfysize=3in
\begin{picture}(300,200)
\put(147,55){$\beta _{coll}$}
\put(182,160){$\beta _{coll}$}
\put(115,75){$\beta _{AdS}^{in}$}
\put(140,120){$-\beta _{AdS}^{out}$}
\put(55,107){contraction}
\put(40,1){\leavevmode\epsfbox{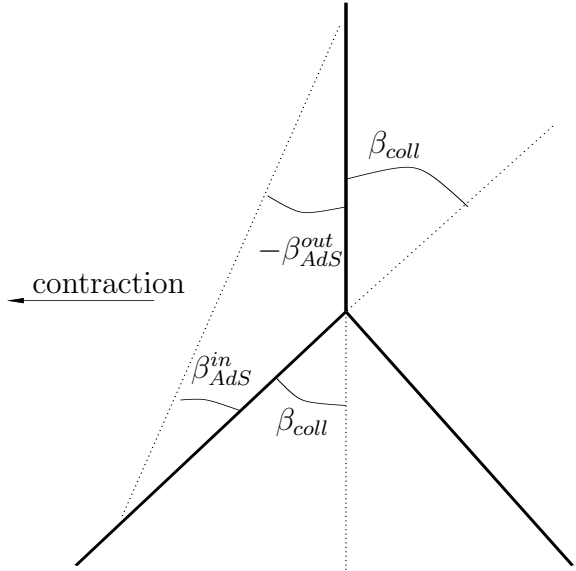
}}
\end{picture}
\end{center}
\caption{{\bf Hyperbolic angles characterizing the brane collision
geometry.} In the exterior $AdS$ a positive angle corresponds to the
direction of an expanding brane. The dotted line in the $AdS$ space
indicates a contour of constant $R,$ or equivalently the worldline of
 an observer at rest with respect to the $AdS$ bulk.}
\label{Fig4}
\end{figure}

The absence of the conical defect at the point of collision implies that
\be
\beta _{AdS}^{out}=\beta _{AdS}^{in}-\beta _{coll},
\label{matching}
\ee
where the hyperbolic angles $\beta _{coll},$ $\beta _{AdS}^{in},$ and $\beta _{AdS}^{out}$
indicate the orientations of the branes with respect to the apparent bulk
spacetime rest frames, which have been cut and pasted together as
shown in Fig.~\ref{Fig4}.
 It follows that the expansion rate on the final brane
just subsequent to the collision is given by
\begin{eqnarray}
H_{out}&=&{1\over \ell }\sinh [\beta _{AdS}^{out}]~\tanh [\chi _{coll}/\ell ]=
        {1\over \ell }\sinh [\beta _{AdS}^{in}-\beta _{coll}]~\tanh [\chi _{coll}/\ell ]\cr
&=&
H_{in}\Biggl[ \cosh [\beta _{coll}]-\sinh [\beta _{coll}]~\coth [\beta _{AdS}^{in}] \Biggr] .
\end{eqnarray}
For $M^5$ initially between the branes, one has 
\be
H_{out}=
{\cosh ^2[\beta _{coll}]\over R_{coll}}-{\sinh [\beta _{coll}]\over \ell }
\sqrt{1+{\ell ^2\over R^2_{coll}}\sinh ^2[\beta _{coll}]}.
\ee
For a bubble of critical radius $a,$ we may re-express $H_{out}$ in the form
\be
H_{out}={1\over R_{coll}}+{{R_{coll}}\over a}\left( {1\over a}-\sqrt{{1\over a^2}+{1\over \ell ^2}}\right) .
\label{m5colapse}
\ee 
We observe that since the term in parenthesis is
always negative, the expansion rate for the final brane is 
negative for sufficiently large $R_{coll}$. 

We now generalize to the case where the intervening spacetime is $dS^5,$
having a curvature radius $\ell _{dS},$ rather than $M^5.$ In this case, the 
region initially between the branes is described by the metric
\be
ds^2=-dt^2+\cosh ^2[t/\ell _{dS}]d\chi ^2+\ell _{dS}^2\sinh ^2[t/\ell _{dS}]dH_{(3)}^2,
\label{dS5}
\ee 
and it follows that 
\be
H_{in}={1\over \ell _{dS}}\cosh [\beta _{coll}]~\coth [t/\ell _{dS}]
={\cosh [\beta _{coll}]\over R_{coll}}
\sqrt{1+{R_{coll}^2\over \ell _{dS}^2}}.
\ee 
For $dS^5$ initially between the branes, one has
\be
H_{out}={\cosh ^2[\beta _{coll}]\over R_{coll}}\sqrt{1+{R_{coll}^2\over \ell _{dS}^2}}
-{\sinh [\beta _{coll}]\over \ell }
\sqrt{1+{\ell ^2\over R_{coll}^2}\sinh ^2[\beta _{coll}]+
{\ell ^2\over \ell _{dS}^2}\cosh ^2[\beta _{coll}]}. 
\ee

For an expanding bubble, $R$ and $\beta $ are not independent. Their interrelation 
is readily described by embedding a bubble trajectory having the geometry
of $dS^4$ and a curvature radius equal to $a=\ell _{dS}\sin [\sigma ]$ in $dS^5$ of curvature
radius $\ell _{dS}$ (For details see \cite{jjmb1}). We construct $dS^5$ 
through the following embedding in (5+1) dimensional Minkowski space:
\begin{eqnarray}
T&=&\ell _{dS}~\sinh [t/\ell _{dS}]~\cosh [\xi ],\cr 
U&=&\ell _{dS}~\cosh [t/\ell _{dS}]~\cos [\chi /\ell _{dS}],\cr 
V&=&\ell _{dS}~\cosh [t/\ell _{dS}]~\sin [\chi /\ell _{dS}],\cr 
X&=&\ell _{dS}~\sinh [t/\ell _{dS}]~\sinh [\xi ]~n_x,\cr 
Y&=&\ell _{dS}~\sinh [t/\ell _{dS}]~\sinh [\xi ]~n_y,\cr 
Z&=&\ell _{dS}~\sinh [t/\ell _{dS}]~\sinh [\xi ]~n_z,
\end{eqnarray}
having the line element previously given in eqn.~(\ref{dS5}), where
$n_x,$ $n_y,$ $n_z$ are directional cosines. $dS^4$ is embedded 
as the surface 
\be
U=\ell _{dS} \cosh [t/\ell _{dS}]~\cos [\chi /\ell _{dS}]=\ell _{dS} \cos [\sigma ]=({\rm constant }).
\ee
It follows that the velocity with respect to the rest frame of the metric
in eqn.~(\ref{dS5}) is 
\be
v=\tanh [\beta _{coll}]=\cosh [t/\ell _{dS}]~{d\chi \over dt}=\cosh [t/\ell _{dS}]
{\tanh [t/\ell _{dS}]\over \tan [\chi /\ell _{dS}]}=
{R_{coll}\cos [\sigma ]\over \sqrt{R^2_{coll}+a^2}},
\ee
where $R=\ell _{dS}~\sinh [t/\ell _{dS}]$ so that
\footnote{Note that as $R_{coll} \to \infty$, $\cosh[\beta _{coll}] \to
\csc[\sigma]$, which is finite rather than infinite. Because of the
inflation of the intervening spacetime, colliding bubbles expanding in $dS^5$
have a finite terminal velocity.}
\be 
\cosh [\beta _{coll}]={\sqrt{R_{coll}^2+a^2}\over \sin [\sigma ]
\sqrt{R_{coll}^2+\ell _{dS}^2}},\quad
\sinh [\beta _{coll}]={R_{coll}\cos [\sigma ]\over \sin [\sigma ]\sqrt{R_{coll}^2+\ell _{dS}^2}}.
\ee
It follows that
\be
H_{out}=
{1\over \ell _{dS}}~{(R^2_{coll}+a^2)\over \sin ^2[\sigma ]R_{coll}
\sqrt{R^2_{coll}+\ell _{dS}^2}}-{1\over \ell }~{\cos [\sigma ]\over \sin [\sigma ]}~
{R_{coll}\over \sqrt{R^2_{coll}+\ell _{dS}^2}}\left( 1+{\ell ^2\over a^2}\right) ^{1/2}.
\ee
As $R_{coll}\to \infty ,$
\be
H_{out}\to {1\over \sin ^2[\sigma ]~\ell _{dS}}-
{\cos [\sigma ]\over \sin [\sigma ]}\sqrt{{1\over \ell ^2}
+{1\over a^2}}.
\ee
Consequently, when 
\be
\cos [\sigma ]\left( 1+{a^2\over \ell ^2}\right) ^{1/2}<1,
\ee 
or equivalently when
\be
\left( 1-{a^2\over \ell _{dS}^2}\right) \left( 1+{a^2\over \ell ^2}\right) <1,
\label{condition}
\ee
a situation occurs where the final brane
is always expanding, for arbitrary large radius of curvature $R_{coll}$ 
of the surface of collision.

The condition in eqn.~(\ref{condition})
is equivalent to the condition
\be
a^2>\ell _{dS}^2-\ell ^2. 
\ee

Finally, we remark on what happens when some of our simplifying
assumptions made to facilitate this calculation are relaxed. In
our calculation we assumed that all the energy available in the 
collision is deposited on the final brane. One may ask what happens 
when some of this energy is carried away by other branes or 
by radiation. We may also ask what happens when the branes are
not infinitely thin. We retain 
the assumption of a symmetric collision. 

First, it is possible to show that if energy, or more precisely
momentum, escapes from the collision away from the central final brane, $H_{out}$
becomes smaller or more negative. This can be seen by considering the 
family of geodesics in the midplane of the collision, which
after the collision become the fiducial observers on the 
FRW universe on the final brane. These fiducial observers
can be continued back into the past prior to the collision
into the initial intervening bulk spacetime, where they
have an initial Hubble constant $H_{before,~center},$ which suffers
a discrete jump at the collision. This jump is proportional
to the impulse in $T_5^5$ at the collision, which in turn 
is proportional to the momentum transfer across the collision. 
More explicitly, the Einstein equations in five dimensions give on the 
4-D plane of symmetry
\be
{\ddot R\over R}+{\dot R^2\over R^2}+{\kappa\over R^2}=-{8\pi G_{(5)}\over 3}T_5^5.
\ee
Here $G_{(5)}$ is the five dimensional Newton constant, $\kappa /R^2$ 
the spatial curvature term, and $T_5^5$
the ``pressure'' along the fifth dimension. 
Assuming a sharp impulse, we obtain
\be
\Bigl( H_{after}-H_{before}\Bigr)_{center} ={(\dot R_{after }-\dot R_{before})\over R}
=-{8\pi G_{(5)}\over 3}\int _{-\epsilon }^{+\epsilon }d\tau ~T_5^5=
-{8\pi G_{(5)}\over 3}I_{5}^5,
\label{jump}
\ee
where $\epsilon$ is taken to be infinitesimal. If follows from 
stress-energy conservation that
\be 
I_5^5=p_{after}^5-p_{before}^5,
\label{conserv}
\ee
where $p^5$ is taken positive in the sense away from the
collision. Eqn.~(\ref{conserv}) may be verified by surrounding the collision by
an infinitesimal pillbox and integrating $T^{\mu 5}$ over its surface.

The integral $I_5^5$ is therefore minimized in the calculation presented
above, where the collision resembles an `inelastic' collision
with nothing bouncing off. If the branes do not completely
stick to each other, $p_{after}^5$ is increased, consequently decreasing
$H_{after}.$ Note that for thick branes the same analysis may be repeated
in the midplane of the brane. 

\section{NUMERICAL CALCULATIONS.}

\begin{figure}
\begin{center}
%\epsfxsize=3.5in
%\epsfysize=4.5in
%\epsfxsize=4.8in
%\epsfysize=3.in
%\begin{picture}(300,200)
%\put(130,210){\Large{$V[\phi]$}}
%\put(310,100){\Large{$\phi$}}
%\put(33,10){\Large{$AdS_1$}}
%\put(210,10){\Large{$AdS_2$}}
%\put(0,-100){\leavevmode\epsfbox{potential-toy.eps}}
%\put(-30,-10){\leavevmode\epsfbox{Fig5.eps}}

\epsfxsize=4.4in
\epsfysize=2.5in
\begin{picture}(300,200)
\put(130,210){\Large{$V[\phi]$}}
\put(310,100){\Large{$\phi$}}
\put(33,10){\Large{$AdS_1$}}
\put(210,10){\Large{$AdS_2$}}
\put(-15,20){\leavevmode\epsfbox{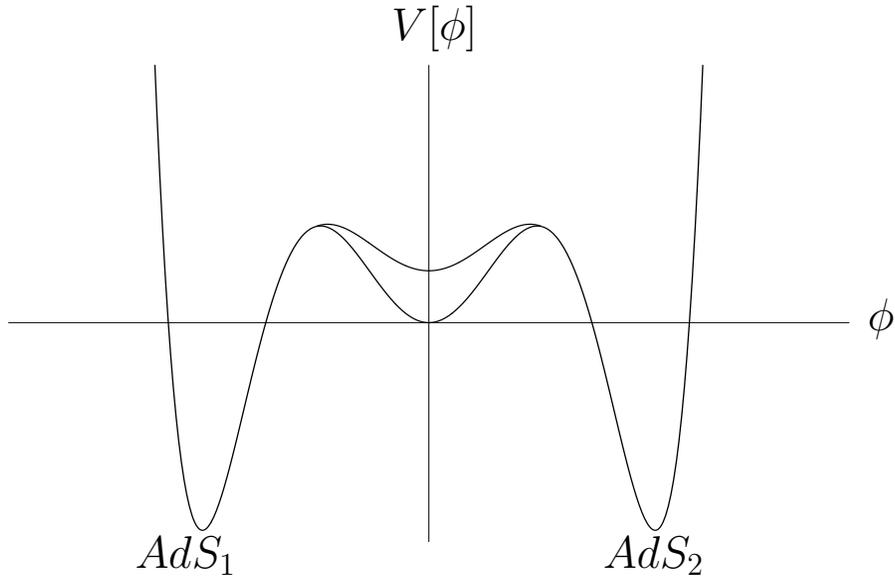}}
\end{picture}
\end{center}
\caption{{\bf Scalar Field Potential}. The initial state for the scalar
field is a metastable  minimum at $\phi=0$ where the spacetime geometry is
either $M^5$ or $dS^5$. The field decays to either of the
two AdS true vacua.}  
\label{Fig5}
\end{figure}

\begin{figure}
\begin{center}
%\epsfxsize=4.in
%\epsfysize=3.in
%\begin{picture}(200,200)
%\put(60,180){\Large{Region I}}
%\put(200,40){\Large{Region II}}
%\put(-60,-10){\leavevmode\epsfbox{Fig6.eps}}
\epsfxsize=4.in
\epsfysize=2.5in
\begin{picture}(200,200)
\put(80,180){\Large{Region I}}
\put(250,40){\Large{Region II}}
\put(-40,5){\leavevmode\epsfbox{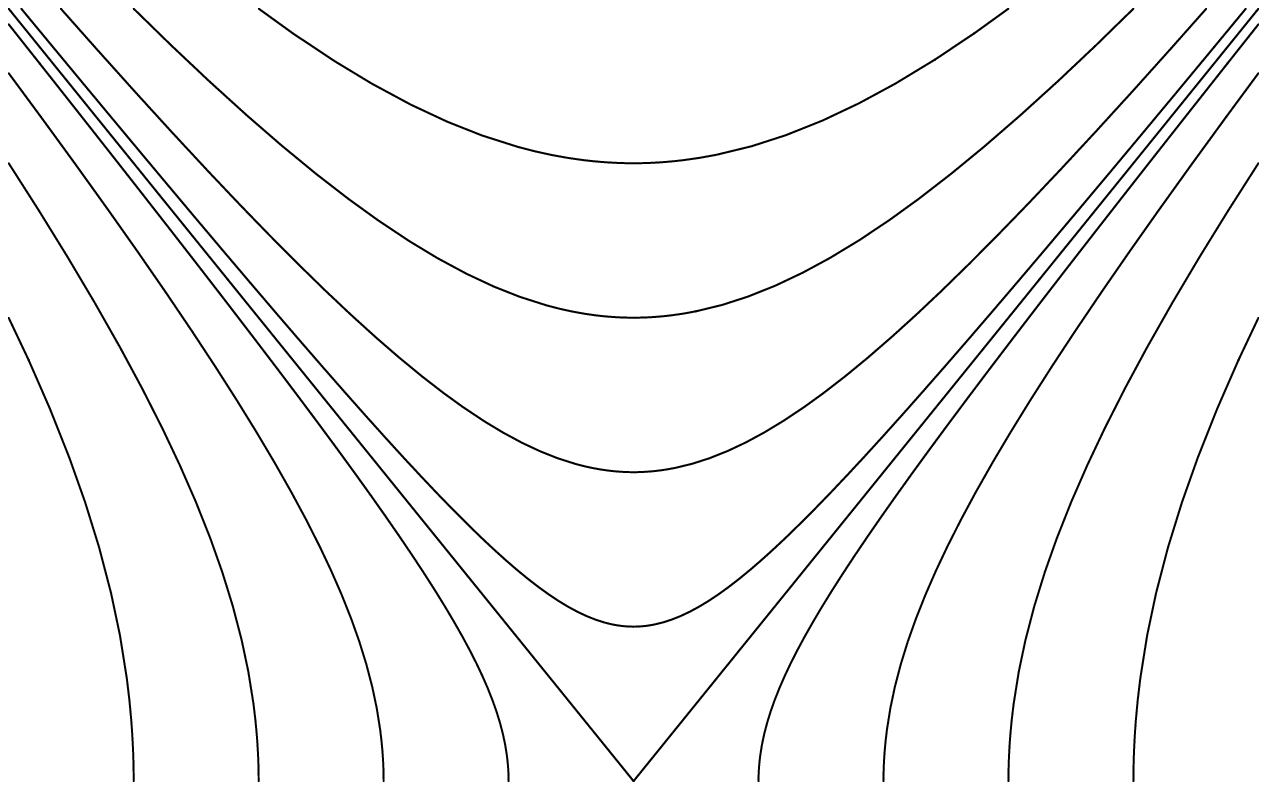}}
\end{picture}
\end{center}
\caption{{\bf Spacetime diagram of the bubble nucleation}. Region I, in the
interior of the bubble is an open FRW universe. Region II is the
exterior of the bubble. The scalar field $\phi$ is constant along the
lines shown in this picture.}  
\label{Fig6}
\end{figure}

We now present some numerical simulations of collisions of expanding bubbles
to explore what occurs in the thick-wall case. 
To study the gravitational effects of the bubble collision
we will concentrate on a model with a single scalar field coupled to
gravity described by the action,\footnote{
We set $8\pi G_{(5)}=1$ unless otherwise stated.}
\be
S=\int d^5x\sqrt{-g}\left[{\frac{1}{2}}R^{(5)}-\frac{1}{2}\partial _M \phi 
\partial ^M \phi -V[\phi ]\right]
\ee
where the self interacting potential for the scalar field is of
the type drawn in Fig.~\ref{Fig5}. This simple potential has the correct 
properties to describe the colliding bubble braneworld scenario.
We assume an initial state where the scalar field is trapped in
a metastable state at $\phi=0$, where cosmological constant is either
positive or zero so that the bulk spacetime outside the bubbles is
either de Sitter or Minkowski space, respectively.
The field tunnels to the true vacuum by the nucleation of bubbles filled with
anti de Sitter space. 
We first discuss the situation with only one bubble, so that
the symmetry of the problem is $SO(4,1)$ and the equations for
the instanton describing the decay of the false vacuum are 
\be
\phi''+4\left(\frac{b'}{b}\right)~\phi' =+V_{,\phi }
\label{inst-phi}
\ee
and 
\be
\left(\frac{b'}{b}\right)^2 =\frac{1}{6} \left[\frac{1}{2}(\phi')^2
-V[\phi ]\right]+\frac{1}{b^2},
\label{inst-b}
\ee
where the Euclidean metric of the instanton is 
\be
d\bar s_E^2= d\sigma ^2+b^2(\sigma )\left[d\tilde t_E^2+\cos^2[\tilde t_E]d\Omega _{(3)}^2\right]
%d\rho^2+b^2(\rho)\left[ dt_E ^2+\cos ^2[t_E]d\Omega _{(3)}^2\right]
\ee
and the prime denotes differentiation with respect to $\sigma$.
Upon analytic continuation of $\tilde t_E$ to $\tilde t =i\tilde t_E,$ the line element becomes
\be
d\bar s_L^2=d\sigma ^2+b^2(\sigma )\left[ -d\tilde t^2+\cosh ^2[\tilde t]d\Omega _{(3)}^2\right],
%d\bar s_L^2=d\rho^2+b^2(\rho)\left[ -d\tau ^2+\cosh ^2[\tau ]d\Omega _{(3)}^2\right] ,
\ee
describing the instanton solution in region II of Fig.~\ref{Fig6}.  

This solution may be continued into region I (i.e., the interior of the 
forward lightcone of the nucleation center) by solving the coupled
equations
\be
\ddot\phi +4\left(\frac{\dot a}{a}\right)~\dot\phi  =-V_{,\phi }
\label{lorz-phi}
\ee
and 
\be
\left(\frac{\dot a}{a}\right)^2 =\frac{1}{6} \left[\frac{1}{2}\dot \phi ^2
+V[\phi ]\right] +\frac{1}{a^2}
\ee
with line element
\be
d\bar s^2=-d\tau^2+a^2(\tau)\left[ d\chi ^2+\sinh ^2[\chi ]d\Omega _{(3)}^2\right] .
\label{lorz-b}
\ee
This system of equations describes the interior of the bubble as an
open (hyperbolic) universe filled with a scalar field $\phi$ and evolving with a
scale factor given by $a(\tau)$\cite{deluccia}.
Due to the negative value of the cosmological constant inside of the bubble,
the initial expansion of this open universe slows down and eventually 
reverses, leading to a collapsing phase. 
In the thin-wall limit, $\phi =\phi _{tv}$
exactly in region I and the geometry is precisely
that of perfect anti de Sitter space, with $a(\tau)=H^{-1}\sin [H\tau].$ In this
idealized case, $\tau=\pi H^{-1}$ corresponds to a coordinate singularity. 
However, with any non-vanishing bubble wall thickness, the wall 
always has a decaying tail, so that
at the lightcone $\phi $ is always slightly displaced from its true vacuum
value.
For the case of an expanding interior geometry, the deviation from 
$\phi _{tv}$ oscillates and decays, with $\phi $ becoming increasingly
close to $\phi _{tv}$ as one passes farther into the bubble interior.
However, when the expansion slows down eventually turning into a collapse,
the deviation from $\phi _{tv}$ has two modes: one regular and the other divergent 
as the ``Big Crunch" of the collapse is approached. The ``Big Crunch"
is a backward 
directed lightcone when the gravitational backreaction of these modes is ignored.
Unless the potential is finely tuned with infinite precision, at least
some amplitude of the divergent mode 
is always present. This divergent mode leads to a spacelike coordinate singularity
\cite{deluccia,abbott}, akin to the 
initial singularity of a homogeneous and isotropic universe with a 
hyperbolic spatial geometry. 

The singularity forms as the result of the extremely finely tuned 
focusing of the scalar field toward the same point
in a large region of spacetime.
The role of the negative cosmological constant is subsidiary. It 
is important only during the middle phase of the evolution between
the two lightcones, where it acts as a sort of perfect lens without
any aberration at all.
Once the initially divergent field mode has been refocused, the negative
cosmological constant is no longer needed. The formation of the 
singularity would occur, even in Minkowski space, for example,
where all but the last moments of the collapse are described by
a sort of upside down Milne universe.  
In the presence of sufficiently large quantum fluctuations exceeding
a certain threshold, away from the semi-classical $(\hbar \to 0)$
limit, the degraded quality of the focusing presumably allows the formation
of a singularity to be avoided.
 
The original motivation for carrying out these simulations
was to study this singularity in the presence of a  collision
with another bubble and its 
consequences for the colliding braneworld scenario (See \cite{jjmb2}).
The argument was that the collision of scalar field bubbles would most
likely produce some radiation which would propagate in the background
spacetime described above. This radiation could however 
spoil the focusing of a certain part of the singularity, thus
opening a window into infinite AdS space. This is what we set out to
study by doing numerical simulations of the colliding bubbles in the
scalar field model described above. 
The simulations performed, however, indicate the presence of another
singularity forming in the future of the collision region. This
implies that the 4d universe within the plane of symmetry that would
be our braneworld would rapidly collapse. These numerical results led
us to investigate this issue further and this paper reports on this
line of work.

\subsection{Fixed Minkowski background}

In order to gain a better understanding of what to expect from the
simulation of the bubble collision we first simulated 
a collision with the background fixed to Minkowski space.
We first find the solution for the scalar field from the instanton 
equation (\ref{inst-phi}), fixing the background metric to be
Minkowski, in other words with $b(\sigma)=\sigma$. We then use this 
solution to generate the initial state with two bubbles, by just
pasting together two of these solutions at some distance $d$ and reversing the
sign of $\phi$ for one of them.
We expect this initial data to
be a good approximation to the real situation of the nucleation of
two bubbles at the same time at sufficiently large separation $d$.
To evolve this forward in time, we make use of the symmetry of the
problem by choosing the hyperbolic slicing of Minkowski space given by
\ba
d\bar s^2= -dt^2+dx^2+t^2~dH^2_{(3)},
\label{5DMinMilne}
\ea
where $dH^2_{(3)}=d\xi ^2+\sinh [\xi ]d\Omega _{(2)}^2.$
This is just Minkowski space expressed as the direct
 product of a Milne universe in one dimension less with the line 
$-\infty <x<+\infty $ of a spatial character. It is along this
direction $x$ where the two bubbles are separated by a distance $d$
initially. In this coordinate system the equation of motion for the
scalar field 
becomes
\ba
{{\partial^2\phi}\over{\partial
t^2}}+{{3}\over{t}} {{\partial\phi}\over{\partial t}}-
{{\partial^2\phi}\over{\partial x^2}}= -V_{,\phi }.
\ea
Note that even though this coordinate system does not cover
the whole spacetime, it allows us to study the region of the collision 
between the two bubbles.

\begin{figure}
\begin{center}
\epsfxsize=4.5in
\epsfysize=4.in
\begin{picture}(200,200)
\put(-70,-10){\leavevmode\epsfbox{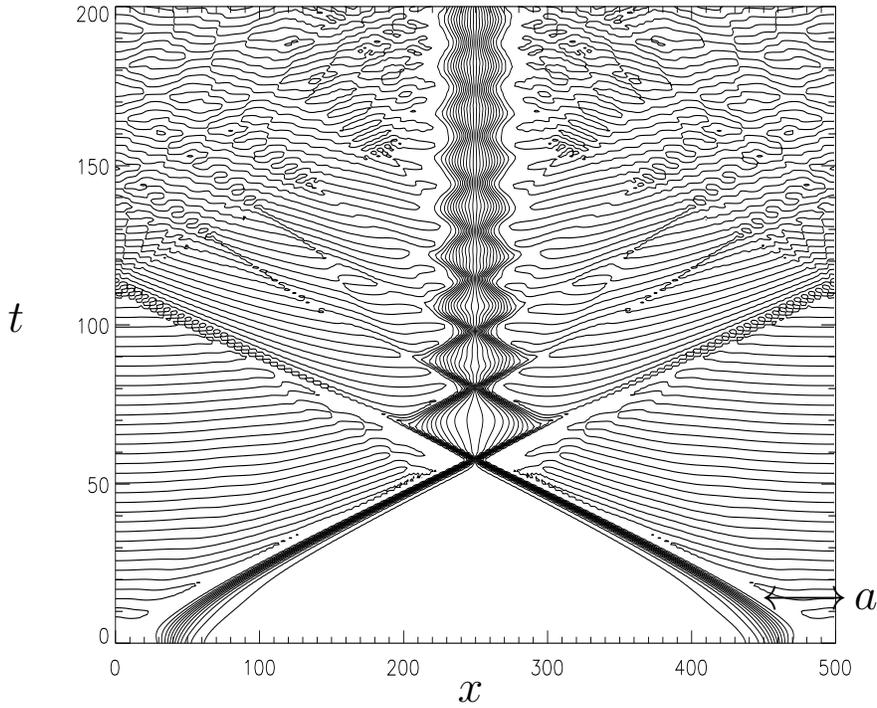}}
\put(100,0){\Large{$x$}}
\put(230,35){\Large{$\rightarrow$}}
\put(215,35){\Large{$\leftarrow$}}
\put(250,35){\Large{$a$}}
\put(-70,140){\Large{$t$}}
\end{picture}
\end{center}
\caption{{\bf Isosurfaces of constant scalar field $\phi$ for a collision
of two bubbles in the $(x,t)$ plane of the 5D-Minkowski space}. The
bubbles collide and bounce off each other emitting radiation at
each encounter till finally joining to form a thick brane that 
interpolates between the two AdS minima of the scalar potential.}  
\label{Fig7}
\end{figure}

In Fig.~\ref{Fig7} we plot the isosurfaces of constant scalar field in the
$(x,t)$ plane. Each point in this 2D picture represents a hyperboloid
of radius $t$ on which the scalar field $\phi$ is constant. We see from this
simulation that the bubbles start at rest with an initial radius $a$
and rapidly accelerate toward one another
until finally colliding with a considerable amount of kinetic
energy at the plane of symmetry $x=0$. At the collision the two
bubbles bounce off one another while continuing to expand but the attractive
force between them bring them together again and again until they
finally settle down to a stable configuration of a kind of ``thick''
domain wall or brane. This is the braneworld where we live. 

One important point here is boundary conditions. We have imposed symmetric
boundary conditions for the field $\phi$ and this allows for the
radiation to be reflected on the boundaries and enter the region of
interest again. This is not desirable and we will see how we do not
have to impose any boundary conditions of this type when we include
gravity.

\subsection{Including gravity}

We now study the same problem including gravity.
As before, it is most convenient to employ a system of coordinates that
maximally exploits the $SO(3,1)$ symmetry of the problem.
To cover all the spacetime in this way would require several coordinate patches,
separated by boundaries across which the geometry of the orbit of a point 
under the action of $SO(3,1)$ changes from $H_{(3)}$ to $dS_{(3)}.$ However, for
our purposes, because our primary interest is the region of collision
and the brane between the bubbles arising from this region, it suffices
to consider one such patch, described by a metric of the form
\ba 
d\bar s^2= -f^2(u,v)~du~dv+r^2(u,v)~dH^2_{(3)}.
\ea
To further fix the gauge, let us additionally restrict the coordinate patch 
to the region $(u+v)>0$ and require that $r=0$ on the line $(u+v)=0.$ The 
gauge may be completely fixed by requiring that $f^2(u,-u)=1.$ In this gauge,
we may rename $\sigma _+=u,$ $\sigma _-=-v,$ so that the metric becomes
\ba 
d\bar s^2= +f^2(\sigma _-,\sigma _+)~d\sigma _+~d\sigma _-+r^2(\sigma _-,\sigma _+)~dH^2_{(3)}
\ea
where, physically, the coordinates $\sigma _+$ and $\sigma _-$ of a point $P$ 
indicate the intersections of the right-moving and left-moving null geodesics
passing through $P$ with the $r=0$ line, where $\sigma $ represents the
physical distance along this line.

We now show how to describe the general expanding bubble solution, endowed
with an $SO(4,1)$ symmetry in terms of the $SO(3,1)$ symmetric coordinates just described. 
In terms of the natural coordinates of the larger $SO(4,1)$ symmetry, the 
solution is described using two coordinate patches, a region I (inside
the forward lightcone of the nucleation center), with the metric
\ba
d\bar s^2&=& -d\tau ^2+a^2(\tau )~dH_{(4)}^2\nonumber\\
         &=& -d\tau ^2+a^2(\tau )~\left[ d\chi ^2+\sinh ^2[\chi ]d\Omega _{(3)}^2\right] \nonumber\\
         &=& -d\tau ^2+a^2(\tau )~\left[ d\xi ^2+\cosh ^2[\xi ]dH _{(3)}^2\right] ,
\label{mt:in}
\ea
and a region II, outside the lightcone, with the metric 
\ba
d\bar s^2&=& d\sigma ^2+b^2(\sigma )~d(dS)_{(4)}^2\nonumber\\
         &=& d\sigma ^2+b^2(\sigma )~\left[ -d\tilde t^2+\cosh ^2[\tilde t]d\Omega _{(3)}^2\right] \nonumber\\
         &=& d\sigma ^2+b^2(\sigma )~\left[ -d\hat t^2+\sinh ^2[\hat t] dH _{(3)}^2\right] .
\label{mt:out}
\ea
The coordinates on the last line of the previous equations cover only
a subregion of the region covered by the coordinates of the preceding line; however, for
studying bubble collisions this does not pose a problem. 

To treat both regions within a common framework, it is useful to transform to null 
coordinates, defining the transformed variables
\ba
t=\ln [\tau ]+\int _0^\tau d\tau \left[ \frac{1}{a(\tau )} - \frac{1}{\tau }\right] 
 =\ln [\tau ]+g(\tau ),
\label{t_def}
\ea
and
\ba
s=\ln [\sigma ]+\int _0^\sigma d\sigma \left[ \frac{1}{b(\sigma )} - \frac{1}{\sigma }\right] 
 =\ln [\sigma ]+f(\sigma ),
\label{s_def}
\ea
so that eqns.~(\ref{mt:in}) and (\ref{mt:out}) become
\ba
d\bar s^2&=& a^2(\tau )\left[ -d(t+\xi )d(t-\xi )+\cosh ^2[\xi ]dH _{(3)}^2\right] \nonumber\\
         &=& a^2(t_{av})\left[ -dt_+dt_-~+\cosh ^2[\xi ]dH _{(3)}^2\right] \nonumber\\
         &=& a^2\left(\frac{t_++t_-}{2}\right) 
         \left[ -dt_+dt_-~+\cosh ^2\left[ \frac{t_+-t_-}{2}\right ] dH _{(3)}^2\right] ,
\ea
and 
\ba
d\bar s^2&=& b^2(\sigma )\left[ d(s+\hat t ~)d(s-\hat t~)+\sinh ^2[\hat t]dH _{(3)}^2\right] 
                                  \nonumber\\
         &=& b^2\left( \frac{s_++s_-}{2}\right) \left[ -ds_+ds_-+\sinh ^2
               \left[ \frac{s_+-s_-}{2}\right] dH _{(3)}^2\right] ,
\ea
respectively.
To patch together the two regions, it is convenient to rescale the null variables back
to $\tau $ and $\sigma ,$ so that the metric is 
\ba 
d\bar s^2=a^2(\tau _{av})
\left[
\frac{-d\tau _-d\tau _+}{a(\tau _{+})a(\tau _{-})}
 +\cosh ^2\left[ \frac{t_+-t_-}{2}\right] dH_{(3)}^2 \right] 
\ea
and
\ba
d\bar s^2=
b^2(\sigma _{av})
\left[
\frac{d\sigma _-d\sigma _+}{b(\sigma _{+})b(\sigma _{-})} +\sinh ^2\left[ \frac{s_+-s_-}{2}\right] dH_{(3)}^2 
\right] .
\ea
Note that the surface of the lightcone corresponds to the limit 
$\tau _+=$(fixed), $\tau _-\to 0+,$
and $\sigma _+=$(fixed), $\sigma _-\to 0+$ in the respective regions inside and outside
the lightcone. Consequently, the lightcone is parameterized equally well by both 
$\sigma _+$ and $\tau _+.$ The correspondence between these variables is established
by matching the curvature radius of the $H_{(3)}$ dimensions by setting
\ba
\lim _{\tau _-\to 0+} 
a^2(\tau _{av})~\cosh ^2\left[ \frac{t_+-t_-}{2}\right]
=\lim _{\sigma _-\to 0+}b^2(\sigma _{av})~
\sinh ^2\left[ \frac{s_+-s_-}{2}\right] .
\ea
From eqn.~(\ref{t_def}), it follows that 
\ba
\tau _{av}e^{g(\tau _{av})}=e^{t_{av}}=e^{(t_++t_-)/2}
=\sqrt{\tau _+\tau _-}e^{[g(\tau _+)+g(\tau _-)]/2}.
\ea
However, as $\tau _-\to 0,$ $g(\tau _-)\to 0,$ and
\ba 
\tau _{av}=\sqrt{\tau _+\tau _-}e^{g(\tau _+)/2},
\ea
and since $a(\tau )=\tau +O(\tau ^3)$ (for else the curvature on the lightcone
would be singular), it follows that
\ba 
a(\tau _{av})=\sqrt{\tau _+\tau _-}e^{g(\tau _+)/2}.
\ea
Similarly, in the same limit
\ba
\cosh ^2\left[ \frac{t_+-t_-}{2}\right] \approx \exp (t_+-t_-)\approx 
\frac{\tau _+}{\tau _-}e^{g(\tau _+)}.
\ea
Hence 
\ba
\lim _{\tau _-\to 0+}
a^2(\tau _{av})~\cosh ^2\left[ \frac{t_+-t_-}{2}\right] =
\tau _+^2e^{2g(\tau _+)}.
\ea 
Likewise, the limit on the right-hand side is equal to $\sigma  _+^2e^{2f(\sigma _+)}.$
It follows that 
\ba 
\tau _+e^{g(\tau _+)}=\sigma  _+e^{f(\sigma _+)},
\label{tran:a}
\ea 
or 
\ba 
t_+=s_+.
\label{tran:b}
\ea

By matching in this way, it is possible to include both the bubble interior
and exterior within the same coordinate patch. $\sigma _+$ and $\sigma _-$
are allowed to take both positive and negative values, subject to the restriction
$\sigma _-\le \sigma _+.$ The range $\sigma _-\le 0\le \sigma _+$ covers the interior
of the lightcone. Geometrically, the transformations
(\ref{tran:a}) and (\ref{tran:b}) map the $\sigma $ line into the $\tau $ axis such that
null geodesics link the two axes. 

If we write the metric in the form
\ba
d\bar s^2=f^2(\sigma _-,\sigma_+)d\sigma_-d\sigma _+
+r^2(\sigma _-,\sigma _+)dH^2_{(3)},
\ea
it follows that
\ba
f^2(\sigma _-,\sigma _+)=\left\{
\begin{array}{lr}
\frac{b^2(\sigma _{av})}{b(\sigma _+)b(\sigma _-)}, &0\le \sigma _-\le \sigma _+,\\
\frac{a^2(\tau _{av})}{b(\sigma _+)b(-\sigma _-)}, &\sigma _-\le 0\le \sigma _+,\\
\frac{b^2(\sigma _{av})}{b(-\sigma _+)b(-\sigma _-)}, &\sigma _-\le \sigma _+\le 0,\\
\end{array}
\right.
\ea
and 
\ba
r^2(\sigma _-,\sigma _+)=\left\{
\begin{array}{lr}
b^2(\sigma _{av})~
\sinh ^2\left[ \frac{s(\sigma _+)-s(\sigma _-)}{2}\right],  
&0\le \sigma _-\le \sigma _+,\\
a^2(\tau _{av})~
\cosh ^2\left[ \frac{s(\sigma _+)-s(-\sigma _-)}{2}\right],  
&\sigma _-\le 0\le \sigma _+,\\
b^2(\sigma _{av})~
\sinh ^2\left[ \frac{-s(\sigma _+)-s(-\sigma _-)}{2}\right],  
 &\sigma _-\le \sigma _+\le 0.\\
\end{array}
\right.
\ea
By considering the appropriate limits, it may be verified that both 
$f^2$ and $r^2$ are smooth across the lightcone. 

We are now ready to rewrite the whole spacetime solution of
the bubble nucleation in terms of the null coordinate system
natural to this problem. This will
be crucial in the next section in order to obtain the initial data for
the numerical simulation of the bubble collision in terms of the one
bubble solutions.

\subsubsection{Equations of motion for a doubly null coordinate system.}

In the doubly null coordinate system given by the metric
\ba 
d\bar s^2=-f^2(u,v)~du~dv+r^2(u,v)~dH^2_{(3)},
\ea
we can write Einstein's equations for a minimally coupled scalar field
in the following way
\ba
r_{uu} f - 2 f_u r_u + {2\over{6}} r f \phi_u^2=0,
\label{uv1}
\ea
\ba
r_{vv} f - 2 f_v r_v + {2\over{6}} r f \phi_v^2=0,
\ea
\ba
2rr_{uv}+4r_ur_v-f^2-{2\over{6}}r^2f^2 V=0,
\ea
\ba
2r^{-1}r_{uv} + f^{-1} f_{uv} +r^{-2} r_u r_v - f^{-2}f_u f_v
-{1\over4}f^2 r^{-2}+ {{2f^2}\over{8}}\left({{2\phi_u
\phi_v}\over{f^2}}- V\right)=0,
\ea
and the scalar field equation becomes
\ba
\phi_{uv} +{3\over2} r^{-1}(r_u\phi_v+r_v\phi_u)+{1\over4}f^2 V_{,\phi }=0.
\label{uv2}
\ea
 
\begin{figure}
\begin{center}
\begin{picture}(200,200)
\epsfxsize=5.5in
\epsfysize=4.in
\put(-112,-30){\leavevmode\epsfbox{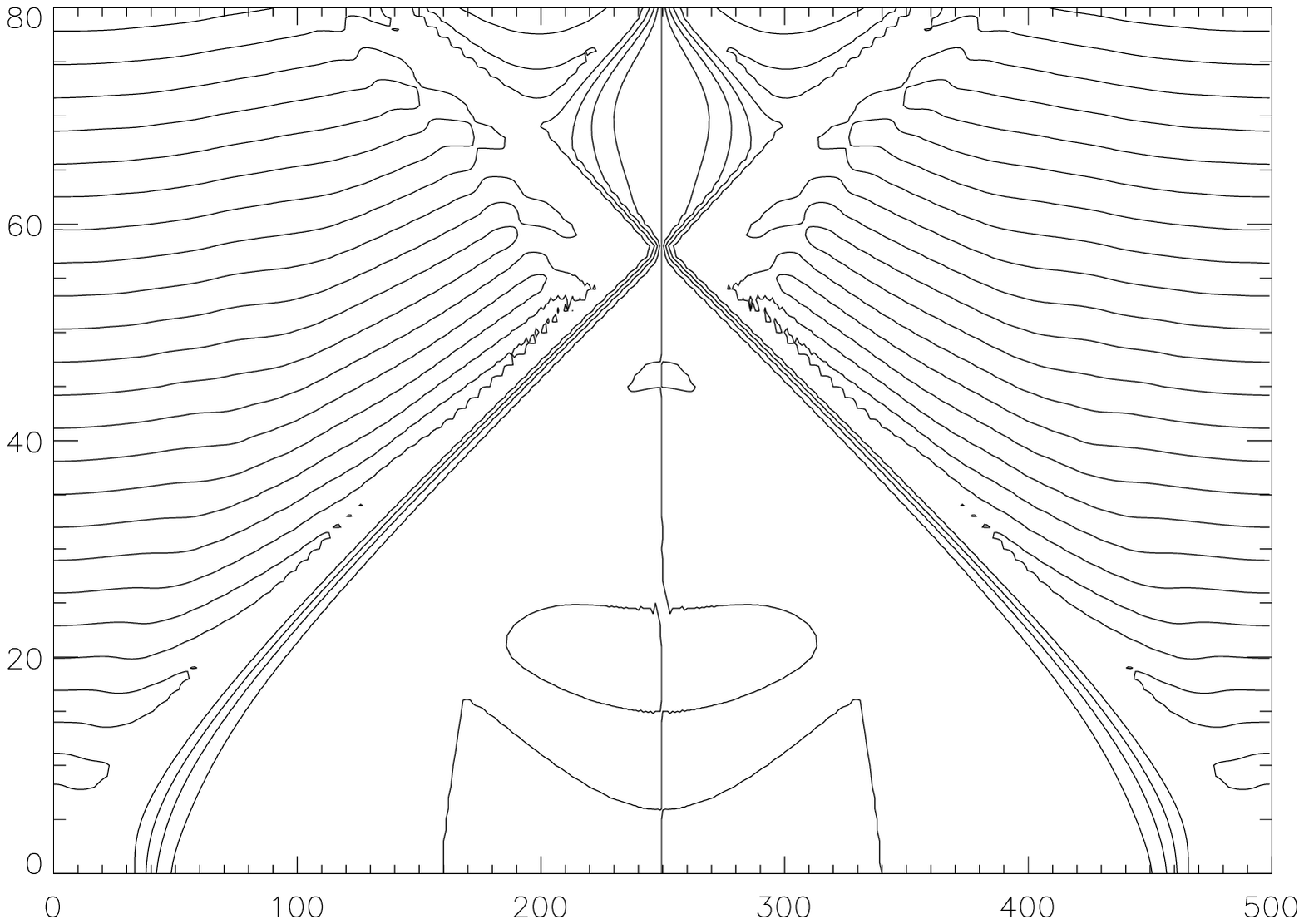}}
\epsfxsize=3.in
\epsfysize=3.in
\put(-5,25){\leavevmode\epsfbox{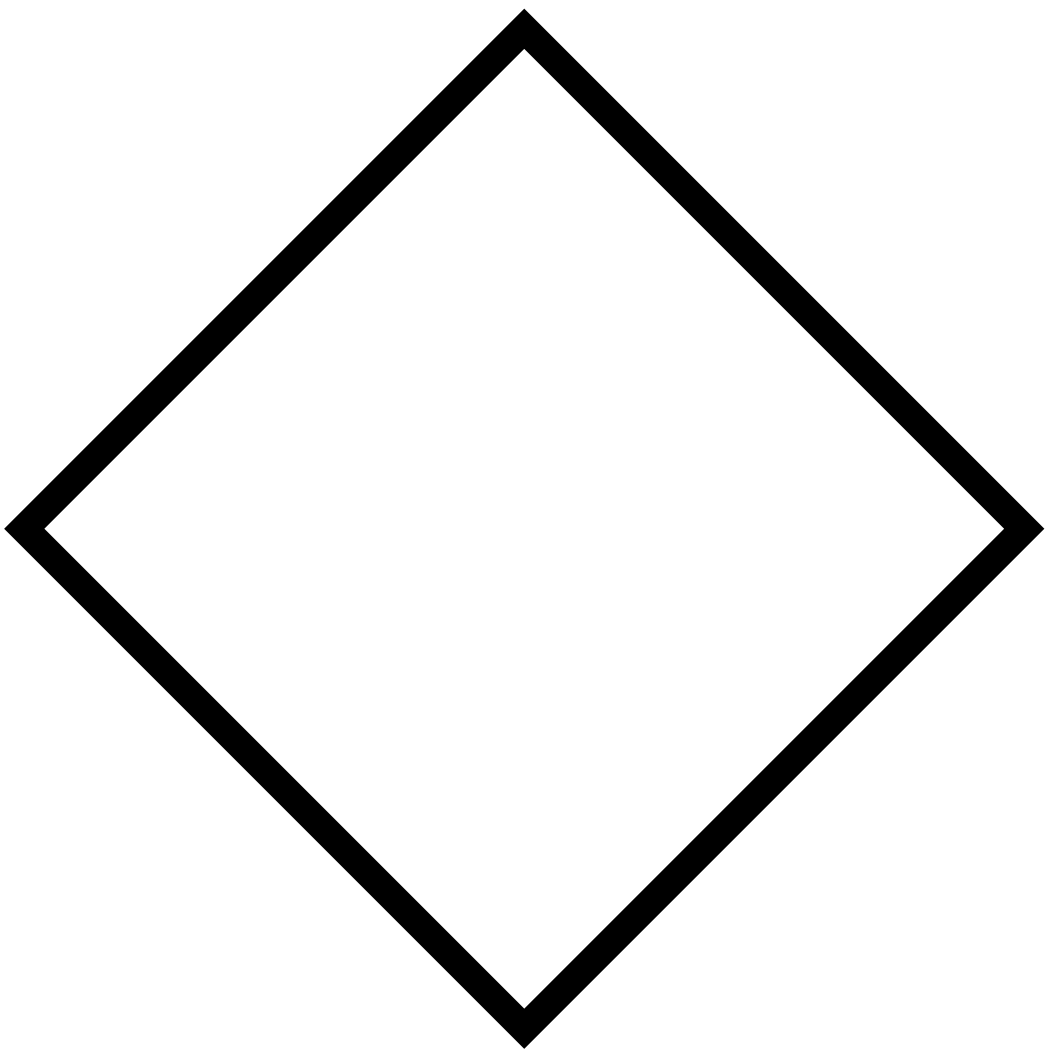}}
\end{picture}
\end{center}
\caption{{\bf Approximate region of the spacetime simulated}. We show the scalar field
solution from the flat background case for reference. It is this region within
the black square that we plot in Figs.~\ref{Fig10} and \ref{Fig11}.}
\label{Fig8}
\end{figure}

In order to integrate these equations of motion numerically, we followed
the techniques developed by Hamade and Stewart\cite{Stewart} with appropriate
 modifications.
We transformed the
problem into the system of first order differential equations 
described in the Appendix. 
We find it convenient to specify initial data on a pair of null surfaces
${\cal N}_L$ and ${\cal N}_R$ at right angles with respect to each other,
chosen so that the bubble collision occurs in the center of the resulting
future diamond. See Fig.~\ref{Fig8}.

\begin{figure}
\begin{center}
\begin{picture}(100,100)
%\put(180,130){\Large{$ u$}}
%\put(-50,130){\Large{$ v$}}
%\put(58,135){\Large{$\cal BW$}}
%\put(125,60){\Large{$\cal N_R$}}
%\put(-10,60){\Large{$\cal N_L$}}
%\put(-140,-140){\leavevmode\epsfbox{initial-data.ps}}
%\put(-140,-40){\leavevmode\epsfbox{Fig9.eps}}
\epsfxsize=4.6in
\epsfysize=2.in
\put(165,125){\Large{$ u$}}
\put(-34,125){\Large{$ v$}}
\put(58,135){\Large{$\cal BW$}}
\put(125,60){\Large{$\cal N_R$}}
\put(-10,60){\Large{$\cal N_L$}}
\put(-95,-20){\leavevmode\epsfbox{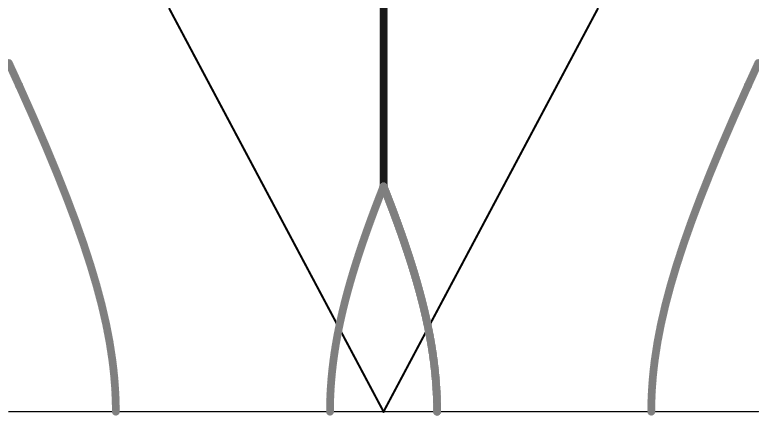}}

\end{picture}
\end{center}
\caption{{\bf Spacetime diagram for the bubble collision}. The initial data
for the simulation is placed on two null surfaces ${\cal N}_L$ and
${\cal N}_R$.} 
\label{Fig9}
\end{figure}

 In our case we use the prescription given in
the previous section to obtain the form of the functions $f(u,v)$,
$r(u,v)$ and $\phi(u,v)$ and their derivatives on the null surfaces ${\cal
N}_{L,R}$ in Fig.~\ref{Fig9} from the known functions of the metric $a(\tau)$
and $b(\sigma)$ as well as from the scalar field in the one bubble
case. We follow the same algorithm as in the fixed Minkowski space
 case and copy the same single bubble data on both branches of $\cal
N$. It is clear from our choice of this surface that this is a good 
approximation if we are mainly interested in what happens around the
collision region, since the existence of the second
bubble would not greatly change the data on this surface.

\section{Numerical checks and Results}

To verify the correctness of our code,
we first simulated the evolution of a single bubble 
to enable comparison to the solution
obtained from eqns.~(\ref{inst-phi})--(\ref{inst-b}) and 
(\ref{lorz-phi})--(\ref{lorz-b}). This is a nontrivial check since
it tests all the technically difficult steps required to go from
the two effectively one-dimensional solutions inside and
outside of the bubble\footnote{We can obtain the
solutions for region I and II separately by solving the one-dimensional
eqns.~(\ref{inst-phi})--(\ref{inst-b}) and
(\ref{lorz-phi})--(\ref{lorz-b}). This is due to the extra 
symmetry present in the one-bubble case.} to its description in terms 
of the new variables in the $(u,v)$ plane. This check shows that the
code based on eqns.~(\ref{uv1})--(\ref{uv2}) reproduces the known solutions for the single
bubble case (even close to the singularity inside of the bubble)
more and more accurately as we increase the fineness of our grid.

\begin{figure}
\begin{center}
\epsfxsize=4.in
\epsfysize=4.in
\begin{picture}(200,200)
\put(100,-10){\Large{$ u$}}
\put(-50,130){\Large{$ v$}}
\put(180,200){\Large{$\cal S$}}
\put(-50,-20){\leavevmode\epsfbox{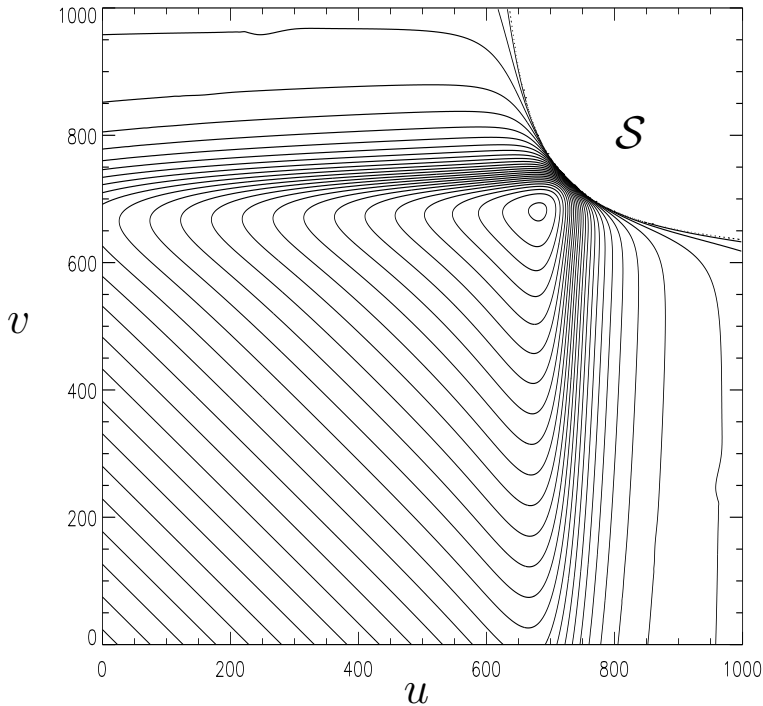}}
\end{picture}
\end{center}
\caption{{\bf Isosurfaces of constant $r(u,v)$ for the collision of 
two bubbles}.}
\label{Fig10}  
\end{figure}

\subsection{Numerical results}

We have simulated the collision of two bubbles with several potentials
and different values of the grid lattice step and the results are
qualitatively the same. The main conclusion is illustrated in Fig.~\ref{Fig10}. 
Here we show contour plots of the function $r(u,v)$ indicating
the {\it size} of the transverse dimensions. We see that 
$r$ increases linearly along the symmetry axis $u=v$
in accord with the Milne-type description of Minkowski space explained
above. This continues to hold until the moment of the collision, after
which $r$ rapidly declines finally vanishing at a singularity.
If we just imagine the freely falling observer following along a $u=v$ geodesic,
she experiences a rapid collapse of the
three dimensional transverse space dimensions finally crashing into
a Big Crunch singularity. 
Fig.~\ref{Fig10} shows that this singularity extends in a 
spacelike manner away from the would-be brane.

\begin{figure}
\begin{center}
\epsfxsize=4.in
\epsfysize=4.in
\begin{picture}(200,200)
\put(100,-10){\Large{$ u$}}
\put(-50,130){\Large{$ v$}}
\put(180,200){\Large{$\cal S$}}
\put(-50,-20){\leavevmode\epsfbox{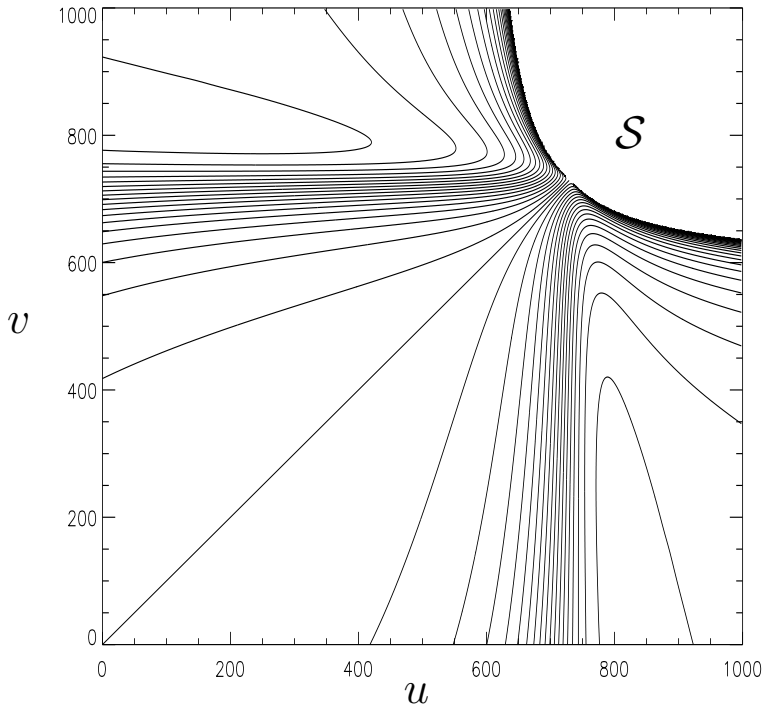}}
\end{picture}
\end{center}
\caption{{\bf Isosurfaces of the scalar field $\phi$.}}  
\label{Fig11}
\end{figure}

We  plot in Fig.~\ref{Fig11} the isosurfaces of constant
$\phi$. The singularity appears here as a divergence in the 
scalar field along the spacelike surface $\cal S$.

These numerical results agree with the analytical results presented
earlier. The source for the  positive pressure in this case is of
course the region around the collision where the gradient for the
scalar field grows from a very small value at the beginning of the
simulation. We have simulated only the collision of bubbles
expanding in Minkowski space.

\section{Discussion}

In this paper we have studied the collisions of $(3+1)$-dimensional
branes embedded in a $(4+1)$-dimensional bulk spacetime from two points
of view: First, we investigated the collisions of thin shells
analytically. Then we studied numerically the collisions of thick
bubbles with gravity included using a two-dimensional doubly null
lattice simulation. We found that in many cases the outcome of the
collision is gravitational collapse to a ``Big Crunch'' singularity on
the final brane, although some situations where this collapse is
averted were found. The thick-wall numerical simulations suggested that
the singularity is not confined to the three transverse dimensions,
but rather extends in a spacelike manner in the fifth dimension.

In a symmetric collision the tendency toward collapse can be
understood as follows. By continuing the fiducial observers of the
final brane into the past, we can formulate a ``jump condition'' at
the bounce, as given in eqn (\ref{jump}). Since for colliding positive energy
branes the impulse  $I_5^5$ is always positive, expansion of the final
brane ($H_{after}>0$) requires a pre-existing expansion in the bulk
(i.e. $H_{before}<0$). This impulse always acts in the sense of
slowing down, if not reversing, the pre-existing bulk expansion. In
some cases, $H_{before}$ is large enough so that $H_{after}$ is
positive. 

\begin{figure}
\begin{center}
\epsfxsize=3in
\epsfysize=3in
\begin{picture}(300,200)
\put(150,50){$M^5$}
\put(70,120){$AdS^5$}
\put(200,120){$AdS^5$}
\put(55,85){contraction}
\put(190,85){contraction}
\put(40,1){\leavevmode\epsfbox{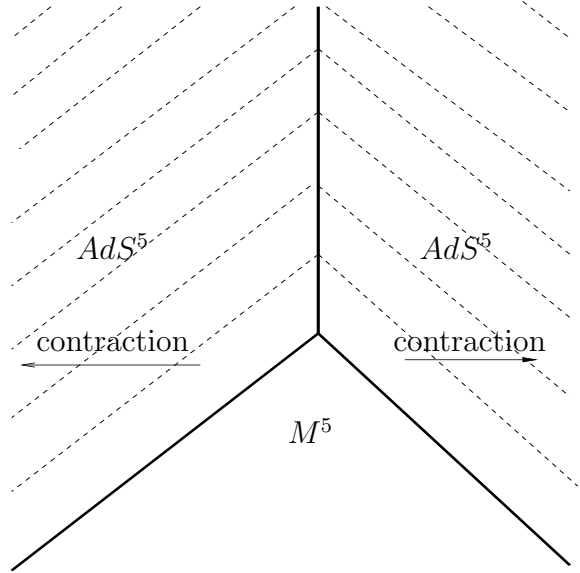
}}
\end{picture}
\end{center}
\caption{{\bf Collision of two positive energy $M^4$ branes initially separated by $M^5$.}
The diagram above indicates a visual representation of why flat branes
with five-dimensional Minkowski space initially intervening always collide to form
a collapsing final brane. The dashed lines indicate contours of constant $R,$ the $R$
of neighboring contours differing by a fixed ratio. Since the branes are of positive
tension, $R$ decreases away from the incident branes. Because $R$ is uniform throughout
the $M^5$ space at the bottom, these contours must be parallel to the incident branes.
Consequently, $R$ on the final brane must decrease in the forward time direction,
indicating collapse.
}
\label{Fig12}
\end{figure}

\begin{figure}
\begin{center}
\epsfxsize=3in
\epsfysize=3in
\begin{picture}(300,200)
\put(130,0){expansion}
\put(63,110){contraction}
\put(185,110){contraction}
\put(40,1){\leavevmode\epsfbox{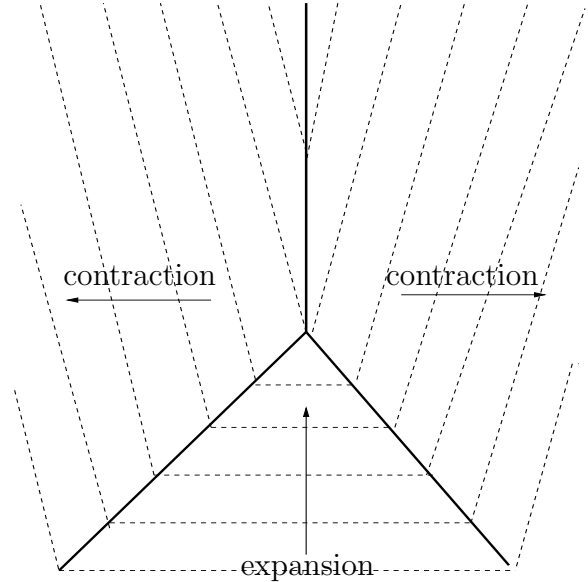
}}
\end{picture}
\end{center}
\caption{{\bf Collision resulting in expansion on the final brane.}
The conclusion displayed in the previous figure that the final
brane is alway collapsing may be circumvented when there is expansion
in the bulk prior to the collision, as shown above. This expansion
may occur either because of a hyperbolic geometry, where $M^5$ has
a Milne geometry, or as the result of a positive bulk cosmological 
constant so that $dS^5$ takes the place of $M^5.$
}
\label{Fig13}
\end{figure}

Our results for the collision of the thin-wall branes may be
demonstrated graphically as in Figs.~\ref{Fig12} and \ref{Fig13}. In
Fig.~\ref{Fig12} we show that two spatially flat colliding branes
initially separated by $M^5$ always produce a collapsing final
brane. The dashed lines indicate contours of constant size $R$ for the
transverse dimensions. This scale is such that $R$ for neighboring
contours differs by a constant factor. In this case, because $R$ is
constant in the $M^5$ initially between the branes, these contours
must be parallel to the incident branes. As seen in the figure, $R$
must contract in the forward time direction along the final brane. In
Fig.~\ref{Fig13} we illustrate how pre-existing expansion in the bulk
can alter this conclusion. Here we show a final brane that is expanding.

We conclude with the following comments:

1. Our analysis supposes the absence of any new physics beyond
   conventional classical five-dimensional Einstein gravity. It is
   possible that new physics could alter the conclusions of our
   analysis. For example, the absence of a deficit angle at the collision,
   as expressed in eqn. (\ref{matching}), can be shown to hold for a singular
   distribution of matter of co-dimension one in conventional
   gravity. New physics or higher order terms, however, could alter this
   conclusion. In the ekpyrotic models where the fifth dimension becomes
   degenerate at the instant of collision, new physics is required and
   our analysis cannot be applied.

2. When the bulk spacetime initially between the colliding branes is
   $M^5$, as indicated in eqn.~(\ref{m5colapse}), whenever $R/a$ is large there is
   always collapse on the final brane. For the Colliding Bubble
   Scenario, small values of $R/a$ are not acceptable because
   $\Omega_{coll}$ would be significantly smaller than unity, leading
   to an empty universe today.

3. For $dS^5$ initially between the branes, if $a^2> l^2_{dS}-l^2$
   there is always expansion for arbitrary large
   $R_{coll}$. Consequently, for $l_{dS} \approx l$ the final brane
   will always be expanding. However, since $a<l_{dS}$,
   $R_{coll}/l_{dS}$ would have to be enormous. In reference
   \cite{ccmb} it was shown that no matter how small the nucleation
   rate is, collisions between large bubbles (with $R_{coll}\gg
   l_{dS}$) would have a strong power law suppression and would
   therefore be extremely unlikely.

\section{Acknowledgments}

We would like to thank 
Oswaldo Dieguez, Jaume Garriga, Marta
 G\'omez-Reino, Chris Gordon, Stephen Hawking, Alberto Alonso
Izquierdo, Sanjay Jhingan, Ken Olum, John Stewart, Neil Turok, 
Alexander Vilenkin, Toby Wiseman and Ivonne 
Zavala for useful conversations. J.J.B-P was supported in
part by the Relativity Group PPARC Rolling Grant. MB was 
supported by Mr Dennis Avery.

\section{Appendix}

\subsection{Equations of motion for the numerical code.}
This Appendix closely follows ref.~\cite{Stewart}.
We implement a first order algorithm to integrate numerically the system of
equations given in Section III for the bubble collision. Using the following definitions
\ba
s=\sqrt{2\over6} \phi,
\qquad
V_c={{2}\over{24}}V,
\qquad
V'_c={{\partial V_c}\over{\partial s}},
\ea
\ba
q=s_v,
\qquad
p=s_u,
\qquad
j=r_u,
\qquad
g=r_v,
\ea
\ba
c={{f_u}\over{f}},
\qquad
d={{f_v}\over{f}},
\ea
\ba
\mu=jq+gp,
\qquad
\lambda=jg-{1\over4}f^2,
\ea
we can rewrite the second order equations of Section III as
\ba
rq_u+{3\over2}\mu +rf^2 V'_c=0,
\qquad
rp_v +{3\over2}\mu +rf^2 V'_c=0,
\ea
\ba
rj_v+2\lambda -2r^2f^2 V_c=0,
\qquad
rg_u+2\lambda -2r^2f^2 V_c=0,
\ea
\ba
r^2c_v-3\lambda +{3\over2}pqr^2+r^2f^2V_c=0,
\qquad
r^2d_u-3\lambda +{3\over2}pqr^2+r^2f^2V_c=0,
\ea

\ba
j_u-2cj+rp^2=0,
\ea
\ba
g_v-2dg+rq^2=0.
\ea
These equations must be supplemented, following the prescription given
in the main text, with initial data on the
null surfaces ${\cal N}_L$ and ${\cal N}_R$.

\end{document}